\documentclass[showpacs,preprintnumbers,amsmath,amssymb,nofootinbib,floatfix]{revtex4}

\usepackage[dvips]{graphicx}
\usepackage{dcolumn}
\usepackage{bm}

\def\mapgeq{\mathbin{\lower.3ex\hbox{$\buildrel>\over{\smash{\scriptstyle\sim}\vphantom{_x}}$}}}
\def\mapleq{\mathbin{\lower.3ex\hbox{$\buildrel<\over{\smash{\scriptstyle\sim}\vphantom{_x}}$}}}
\def\mapgeqeq{\mathbi{\lower.3ex\hbox{$\buildrel>\over{\smash{\scriptstyle\approx}\vphantom{_2}}$}}}
\def\mapleqeq{\mathbin{\lower.3ex\hbox{$\buildrel<\over{\smash{\scriptstyle\approx}\vphantom{_2}}$}}}
\def\Journal#1#2#3#4{{#1} {\bf #2}, #3 (#4)}
\def\MPL{Mod. Phys. Lett. A}

\def\NPB{Nucl. Phys. B}

\def\NPSUPPL{Nucl. Phys. Proc. Suppl.}
\def\PLB{{Phys. Lett.} B}

\def\PLBOLD{Phys. Lett.}
\def\PRL{Phys. Rev. Lett.}
\def\RMP{Rev. Mod. Phys.}
\def\PRD{Phys. Rev. D}

\def\PTP{Prog. Theor. Phys.}
\def\JHEP{JHEP}

\def\EPJ{Euro. Phys. J. C}

\def\JETPUSSR{Sov. Phys. JETP}
\def\ZETP{Zh. Eksp. Teor. Piz.}

\def\IJMP{Int. J. Mod. Phys. A}

\def\JPG{J. Phys. G}

\def\NIMA{Nucl. Instrum. Methods A}
\def\NJP{New. J. Phys.}
\def\Erratum{Erratum-ibid}

\begin{document}

\preprint{TOKAI-HEP/TH-0502}

\title{Neutrino Mass Textures with Maximal CP Violation}

\author{Ichiro Aizawa}
\email{4aspd001@keyaki.cc.u-tokai.ac.jp}

\author{Teruyuki Kitabayashi}
\email{teruyuki@keyaki.cc.u-tokai.ac.jp}

\author{Masaki Yasu\`{e}}%
\email{yasue@keyaki.cc.u-tokai.ac.jp}
\affiliation{\vspace{5mm}%
\sl Department of Physics, Tokai University,\\
1117 Kitakaname, Hiratsuka, Kanagawa 259-1292, Japan\\
}

\date{April, 2005}

\begin{abstract}
We show three types of neutrino mass textures, which give maximal CP-violation as well as maximal atmospheric neutrino mixing.  These textures are described by six real mass parameters: one specified by two complex flavor neutrino masses and two constrained ones and the others specified by three complex flavor neutrino masses.  In each texture, we calculate mixing angles and masses, which are consistent with observed data, as well as Majorana CP phases.
\end{abstract}

\maketitle
\section{\label{sec:1}Introduction}
To discuss whether CP violation in neutrino oscillations \cite{CPphases} exists in Nature is the next research subject in neutrino physics \cite{CPViolation}, whose properties have been clarified by various experiments \cite{experiments,NeutrinoSummary} since the Super-Kamiokande's confirmation of the neutrino oscillations \cite{SK}.  To describe CP violation, the PMNS neutrino mixing matrix \cite{PMNS}, $U_{PMNS}$, includes CP violating phases denoted by $\delta$ and $\beta_{1,2,3}$, which are parameterized in $U_{PMNS}=U_\nu K$ 
with
\begin{eqnarray}
U_\nu&=&\left( \begin{array}{ccc}
  c_{12}c_{13} &  s_{12}c_{13}&  s_{13}e^{-i\delta}\\
  -c_{23}s_{12}-s_{23}c_{12}s_{13}e^{i\delta}
                                 &  c_{23}c_{12}-s_{23}s_{12}s_{13}e^{i\delta}
                                 &  s_{23}c_{13}\\
  s_{23}s_{12}-c_{23}c_{12}s_{13}e^{i\delta}
                                 &  -s_{23}c_{12}-c_{23}s_{12}s_{13}e^{i\delta}
                                 & c_{23}c_{13},\\
\end{array} \right),
\nonumber \\
K &=& {\rm diag}(e^{i\beta_1}, e^{i\beta_2}, e^{i\beta_3}),
\label{Eq:U_nu}
\end{eqnarray}
where $c_{ij}=\cos\theta_{ij}$ and $s_{ij}=\sin\theta_{ij}$ ($i,j$=1,2,3) are the mixing angles for three massive neutrinos of ($\nu_1$, $\nu_2$, $\nu_3$) whose masses are, respectively, denoted by ($m_1$, $m_2$, $m_3$).  The CP violating Majorana phases are specified by two combinations of $\beta_{1,2,3}$ such as $\beta_i-\beta_3$ ($i$=1,2,3).  The mixing angles of $\theta_{12}$, $\theta_{23}$ and $\theta_{13}$ are, respectively, identified with the observed mixing angles of $\theta_\odot$ for the solar neutrino mixing$, \theta_{atm}$ for the atmospheric neutrino mixing  and $\theta_{CHOOZ}$ for the $\nu_e$-$\nu_\tau$ mixing.  The observed results indicate that these mixing angles are constrained to be \cite{NeutrinoSummary}:
\begin{eqnarray}
&&0.70 < \sin^2 2\theta_\odot < 0.95,\quad
0.92 < \sin^22\theta_{atm},\quad
\sin^2\theta_{CHOOZ} <0.05.
\label{Eq:MixingData}
\end{eqnarray}
Furthermore, massive neutrinos satisfy that
\begin{eqnarray}
&&5.4 \times 10^{-5} {\rm eV}^2 < \Delta m_\odot^2 < 9.5 \times 10^{-5} {\rm eV}^2, \quad
1.2 \times 10^{-3} {\rm eV}^2 < \Delta m_{atm}^2 < 4.8 \times 10^{-3} {\rm eV}^2,
\label{Eq:MassData}
\end{eqnarray}
where $\Delta m_\odot^2= m_2^2-m_1^2$ ($>0$) and $\Delta m_{atm}^2= \vert m_3^2-m_2^2\vert$.  It is then widely recognized that the atmospheric and solar neutrino mixings are nearly maximal and that the mass hierarchy of $\Delta m_{atm}^2\gg\Delta m_\odot^2$ exists.  

There have been various discussions to ensure the appearance of the observed bilarge neutrino mixing and mass hierarchy \cite{BilargeSummary} such as those based on the $\mu$-$\tau$ symmetry \cite{Nishiura,mu-tau,mu-tau1,mu-tau2}.  However, the effect of CP violation, namely, the effect of complex flavor neutrino masses in the neutrino mixing itself is not fully understood \cite{CPinMixing}.  If CP violation is included in our discussions, it is not obvious what type of a neutrino mass texture is generally required to explain the observed properties of neutrino oscillations although there have been several specific textures proposed so far \cite{MassTextureCP1,MassTextureCP2,MassTextureCP3}.  Of course, the mixing angles are modified by the inclusion of the imaginary parts of flavor neutrino masses in addition to their ordinary real parts.  In fact, our recent discussions \cite{GeneralCP} have found the novel property that the effect of imaginary parts of flavor neutrino masses is indispensable to describe the atmospheric neutrino mixing.  In other words,  the atmospheric neutrino mixing based on real flavor neutrino masses cannot be determined by the limit of vanishing imaginary parts of complex flavor neutrino masses.

Let us first summarize our results found in Ref.\cite{GeneralCP}.  Our complex flavor neutrino mass matrix of $M_\nu$ is parameterized by
\begin{eqnarray}
&& M_\nu = \left( {\begin{array}{*{20}c}
	M_{ee} & M_{e\mu} & M_{e\tau}  \\
	M_{e\mu} & M_{\mu\mu} & M_{\mu\tau}  \\
	M_{e\tau} & M_{\mu\tau} & M_{\tau\tau}  \\
\end{array}} \right),
\label{Eq:NuMatrixEntries}
\end{eqnarray}
where $U^T_{PMNS}M_\nu U_{PMNS}$=diag.($m_1$, $m_2$, $m_3$).\footnote{It is understood that the charged leptons and neutrinos are rotated, if necessary, to give diagonal charged-current interactions and to define the flavor neutrinos of $\nu_e$, $\nu_\mu$ and $\nu_\tau$.}  Consider a Hermitian matrix of ${\rm\bf M}=M^\dagger_\nu M_\nu$, then, we have obtained that\footnote{This relation is most easily seen by expressing ${\rm\bf M}_{e\mu,e\tau}$ in terms of the neutrino masses, the neutrino mixing angles and the Dirac phase.}
\begin{eqnarray}
&&
\tan \theta _{23}  = \frac{{\rm Im}\left( {\rm\bf M}_{e\mu}\right)}{{\rm Im}\left( {\rm\bf M}_{e\tau}\right)}.
\label{Eq:theta23}
\end{eqnarray}
This relation is only valid if there are complex flavor neutrino masses.  It is the announced property that the atmospheric neutrino mixing angle cannot be determined by Eq.(\ref{Eq:theta23}) in any models without CP violation because no active imaginary parts are present. The CP violating phase of $\delta$ is also expressed by ${\rm\bf M}_{e\mu}$ and ${\rm\bf M}_{e\tau}$ as
\begin{eqnarray}
s_{23} {\rm\bf M}_{e\mu} + c_{23} {\rm\bf M}_{e\tau} = \left| {s_{23} {\rm\bf M}_{e\mu} + c_{23} {\rm\bf M}_{e\tau}} \right| e^{-i\delta }.
\label{Eq:Phase-delta}
\end{eqnarray}
The simple relation arises in the case of maximal CP violation, which is realized by requiring that
\begin{eqnarray}
s_{23} {\rm Re}\left( {\rm\bf M}_{e\mu}\right) + c_{23} {\rm Re}\left( {\rm\bf M}_{e\tau}\right)=0.
\label{Eq:MaximalCPRelation}
\end{eqnarray}
Combining Eqs.(\ref{Eq:Phase-delta}) and (\ref{Eq:MaximalCPRelation}), we find that ${\rm\bf M}_{e\mu,e\tau}$ are related to each other as
\begin{eqnarray}
&&
\sin 2\theta_{23}{\rm\bf M}_{e\tau}=-{\rm\bf M}^\ast_{e\mu}+\cos 2\theta_{23}{\rm\bf M}_{e\mu},
\label{Eq:emu-etau}
\end{eqnarray}
for Since $\delta=\pm \pi/2$.  It should be emphasized that the advantage of the use of ${\rm\bf M}$ lies in the fact that 1) all phases in ${\rm\bf M}$ are related to $\delta$ and 2) phases in $M_\nu$ which are irrelevant for $\delta$ are automatically removed.  However, some of the information of $M_\nu$ including Majorana phases are lost.  

In this article, we examine $M_\nu$ itself suggested by the texture of ${\rm\bf M}$ provided that the atmospheric neutrino mixing and CP violation are both maximal,\footnote{The similar but more general analysis focusing on the quark sector has been given in Ref.\cite{Koide}.} respectively, characterized by $\delta=\pm\pi/2$ and $c_{23}=\sigma s_{23}=1/\sqrt{2}$ ($\sigma =\pm 1$)\footnote{The sign of $s_{23}$ affects the value of $\tan 2\theta_{13}$ via $Y$ in Eq.(\ref{Eq:X-Y}) for a given $M_\nu$ because the change of the sign may cause cancellation in $Y$ giving smaller $\sin^2\theta_{13}$.} and present three types of textures controlled by a single phase of $\theta$.  Two textures with $M_{e\tau}=-\sigma e^{i\theta}M^\ast_{e\mu}$ yield $\tan 2\theta_{12}\propto M_{e\mu}+e^{i\theta}M^\ast_{e\mu}$ and $\tan 2\theta_{13}\propto M_{e\mu}-e^{i\theta}M^\ast_{e\mu}$ while the other texture with $M_{e\tau}=-\sigma e^{i\theta}M_{e\mu}$ gives $\tan 2\theta_{12}\propto 1+e^{i\theta}$ and $\tan 2\theta_{13}\propto 1-e^{i\theta}$.  Majorana CP violation in theses three types of textures, respectively, is absent, active if ${\rm arg}(M_{ee})\neq\theta/2$ and active if ${\rm arg}(M_{ee})\neq{\rm arg}(M_{e\mu})+\theta/2$.

\section{\label{sec:2}Textures with Maximal CP Violation}
Maximal CP violation and maximal atmospheric neutrino mixing arise if the condition of
\begin{eqnarray}
&&
{\rm\bf M}_{e\tau}=-\sigma {\rm\bf M}^\ast_{e\mu}
\label{Eq:MaximalAtm-CP}
\end{eqnarray}
is satisfied in Eq.(\ref{Eq:emu-etau}), which is expressed in terms of the flavor neutrino masses:
\begin{eqnarray}
&&
M^\ast_{ee}M_{e\tau}+M^\ast_{e\mu}M_{\mu\tau}+M^\ast_{e\tau}M_{\tau\tau}
=-\sigma
\left( M_{ee}M^\ast_{e\mu}+M_{e\mu}M^\ast_{\mu\mu}+M_{e\tau}M^\ast_{\mu\tau}\right).
\label{Eq:MaximalAtm-CP-M}
\end{eqnarray}
There are many candidates of $M_\nu$ that satisfy Eq.(\ref{Eq:MaximalAtm-CP-M}) \cite{future}.  Among others, we show the following simpler relations for $M_\nu$ itself with a phase of $\theta$ in $z=e^{i\theta}$:
\begin{itemize}
\item $M_{e\tau}=-\sigma zM^\ast_{e\mu}$ and $M_{\tau\tau}= zM^\ast_{\mu\mu}$ giving
\begin{eqnarray}
&& 
	M^{(1)}_\nu = \left( {\begin{array}{*{20}c}
   M_{ee} & M_{e\mu} & -\sigma zM^\ast_{e\mu} \\
   M_{e\mu} & M_{\mu\mu} & M_{\mu\tau}  \\
   -\sigma zM^\ast_{e\mu} & M_{\mu\tau} & zM^\ast_{\mu\mu}  \\
\end{array}} \right),
\label{Eq:Mnu-1}
\end{eqnarray}
where $M_{ee,\mu\tau}=zM^\ast_{ee,\mu\tau}$ are required,
\item $M_{\mu\tau}=-\sigma M_{ee}$, $M_{e\tau}=-\sigma zM^\ast_{e\mu}$ and $M_{\tau\tau}= zM^\ast_{\mu\mu}$ giving
\begin{eqnarray}
&& 
	M^{(2)}_\nu = \left( {\begin{array}{*{20}c}
   M_{ee} & M_{e\mu} & -\sigma zM^\ast_{e\mu} \\
   M_{e\mu} & M_{\mu\mu} & -\sigma M_{ee}  \\
   -\sigma zM^\ast_{e\mu} & -\sigma M_{ee} & zM^\ast_{\mu\mu}  \\
\end{array}} \right),
\label{Eq:Mnu-2}
\end{eqnarray}
\item $M_{\mu\tau}=-\sigma M_{ee}$, $M_{e\tau}=-\sigma zM_{e\mu}$ , $M_{\mu\mu}=z\kappa M_{e\mu}$ and $M_{\tau\tau}= \kappa^\ast M_{e\mu}$ for an arbitrary complex number $\kappa$ giving
\begin{eqnarray}
&& 
	M^{(3)}_\nu = \left( {\begin{array}{*{20}c}
   M_{ee} & M_{e\mu} & -\sigma zM_{e\mu} \\
   M_{e\mu} & z\kappa M_{e\mu} & -\sigma M_{ee}  \\
   -\sigma zM_{e\mu} & -\sigma M_{ee} &  \kappa^\ast M_{e\mu}  \\
\end{array}} \right).
\label{Eq:Mnu-3}
\end{eqnarray}
\end{itemize}
These textures have six real parameters for a given $z$.  The texture of $M^{(1)}_\nu$ is characterized by two constrained parameters and two complex parameters and those of $M^{(2,3)}_\nu$ are characterized by three complex parameters.  The interesting cases correspond to the simplest choice of $z=\pm 1$ and $z=\pm i$, where $M^{(1)}_\nu$ with $z=1$ yields texture discussed in Ref.\cite{MassTextureCP3,GeneralCP}.

Hereafter, we use the following notations and relations for $\omega$ and $\omega^\prime$ representing general complex variables:
\begin{eqnarray}
&& 
\omega_\pm = \frac{\omega \pm z\omega^\ast}{2},
\quad
\frac{\omega^\prime_+}{\omega_+}=\frac{{\rm Re}\left(\omega^\ast\omega^\prime+z^\ast\omega\omega^\prime\right)}{2\left|\omega_+\right|^2},
\quad
\frac{\omega^\prime_-}{\omega_+}=i\frac{{\rm Im}\left(\omega^\ast\omega^\prime+z^\ast\omega\omega^\prime\right)}{2\left|\omega_+\right|^2},
\nonumber\\
&&
\omega_+ = {\rm Re}\left(\omega, \theta\right)+i{\rm Im}\left(\omega, \theta\right)=\pm e^{i\theta/2}\left| \omega_+\right|,
\quad
\omega_- = {\rm Re}\left(\omega, \theta+\pi\right)+i{\rm Im}\left(\omega, \theta+\pi\right)=\pm ie^{i\theta/2}\left| \omega_-\right|,
\nonumber\\
&& 
{\rm Re}(\omega, \theta)=\frac{\left(1+\cos\theta\right){\rm Re}\left(\omega\right)+\sin\theta{\rm Im}\left(\omega\right)}{2},
\quad
{\rm Im}(\omega, \theta)=\frac{\left(1-\cos\theta\right){\rm Im}\left(\omega\right)+\sin\theta{\rm Re}\left(\omega\right)}{2},
\label{Eq:Notations}
\end{eqnarray}
and
\begin{eqnarray}
&&
\frac{1+z}{z\omega+\omega^\ast}=\frac{2{\rm Re}\left(\left(1+z\right)\omega\right)}{\left| z\omega+\omega^\ast \right|^2},
\quad
\frac{1-z}{z\omega+\omega^\ast}=i\frac{2 {\rm Im}\left(\left(1-z\right)\omega\right)}{\left| z\omega+\omega^\ast \right|^2}.
\label{Eq:Notations2}
\end{eqnarray}
The ratios of particular combinations of $\omega$, $\omega^\prime$ and $z$ turn out to be either real or pure imaginary. It should be noted that ${\rm Re}(\omega, \theta)$ becomes ${\rm Re}(\omega)$ with ${\rm Im}(\omega, \theta)=0$ giving $\omega_+={\rm Re}(\omega )$ for $z=1$ while ${\rm Im}(\omega, \theta)$ becomes ${\rm Im}(\omega)$ and ${\rm Re}(\omega, \theta)=0$ giving $\omega_+=i{\rm Im}(\omega )$ for $z=-1$.  Similarly, for $z=\pm i$, we find the relation of ${\rm Re}(\omega, \theta)=\pm{\rm Im}(\omega, \theta)$ giving $\omega_+=e^{\pm i\pi/4}({\rm Re}(\omega)\pm {\rm Im}(\omega))/\sqrt{2}$.

Once having explicit forms of the neutrino mass textures, we can compute neutrino oscillation parameters as shown in the Appendix \ref{sec:Appendix}. Using the relations found in the Appendix \ref{sec:Appendix}, we obtain results with maximal CP violation ($\delta = \pm\pi/2$) and maximal atmospheric neutrino mixing ($c_{23}=\sigma s_{23}=1/\sqrt{2}$ with $\sigma=\pm 1$).\footnote{In the subsequent discussions, the freedom associated with $M_{\tau\tau}-M_{\mu\mu}$ is not explicitly used.  It can be determined by Eq.(\ref{Eq:theta23-Appendix}) for $\cos 2\theta_{23}=0$.}

For $M^{(1)}_\nu$, parameters used to compute masses from Eq.(\ref{Eq:Masses}) and mixing angles from Eqs.(\ref{Eq:theta12-Appendix})-(\ref{Eq:theta23-Appendix}) are given by
\begin{eqnarray}
&&
X =  \frac{\sqrt 2\left( M_{e\mu} \right)_+}{c_{13}},
\quad
Y = \sqrt 2 \sigma \left( M_{e\mu} \right)_-,
\nonumber\\
&& 
\lambda _1  =  \frac{s_{13}^2}{\cos 2\theta_{13}}
\left(
\left( M_{ee}\right)_+ + \left( M_{\mu\mu} \right)_+ + \sigma \left( M_{\mu\tau}\right)_+
\right) + \left( M_{ee}\right)_+,
\nonumber\\
&& 
\lambda _2  = \left( M_{\mu\mu} \right)_+ - \sigma \left( M_{\mu\tau}\right)_+,
\quad
\lambda _3  = \left( M_{\mu\mu} \right)_+ + \sigma \left( M_{\mu\tau}\right)_+,
\label{Eq:Parameters-1}
\end{eqnarray}
where we have used the property of $M_{ee,\mu\tau}=(M_{ee,\mu\tau})_+$. The mixing angles of $\theta_{12,13}$ are calculated to be:
\begin{eqnarray}
&& 
\tan 2\theta _{12}  =
2\sqrt 2 \frac{\cos 2\theta _{13} \left( M_{e\mu} \right)_+}{c_{13} \left[ \left( 1-3s^2_{13}\right) \left( M_{\mu\mu}\right)_+ - c_{13}^2 \left(\sigma\left(M_{\mu\tau}\right)_++ \left(M_{ee}\right)_+ \right) \right]},
\label{Eq:Angle12-1}\\
&&
\tan 2\theta _{13} e^{i\delta } = 2\sqrt 2 \sigma \frac{\left( M_{e\mu} \right)_-}{\left( M_{\mu\mu} \right)_+ + \sigma \left( M_{\mu\tau}\right)_+ + \left( M_{ee}\right)_+}.
\label{Eq:Angle13-1}
\end{eqnarray}
Because of Eq.(\ref{Eq:Notations}), Eqs.(\ref{Eq:Angle12-1}) and (\ref{Eq:Angle13-1}) give the correct expressions for the mixing angles. As a result, the assumed condition of $e^{i\delta}=\pm i$ is consistent with Eq.(\ref{Eq:Angle13-1}) and the appearance of the required maximal CP violation is assured.
The neutrino masses are calculated to be:
\begin{eqnarray}
&& 
m_1 e^{ - 2i\beta _1 }  = 
\left( M_{\mu\mu} \right)_+-\sigma \left( M_{\mu\tau} \right)_+- \frac{1 + \cos 2\theta _{12} }{\sin 2\theta _{12} }\frac{\sqrt 2 \left( M_{e\mu} \right)_+}{c_{13} },
\nonumber\\
&& 
m_2 e^{ - 2i\beta _2 }  = 
\left( M_{\mu\mu} \right)_+-\sigma \left( M_{\mu\tau} \right)_+ + \frac{1 - \cos 2\theta _{12} }{\sin 2\theta _{12} }\frac{\sqrt 2 \left( M_{e\mu} \right)_+}{c_{13} },
\nonumber\\
&& 
m_3 e^{ - 2i\beta _3 }  = \frac{c^2_{13}\left( \left( M_{\mu\mu} \right)_+ +\sigma \left( M_{\mu\tau} \right)_+\right)+ s^2_{13}\left( M_{ee} \right)_+}{\cos 2\theta _{13}},
\label{Eq:Masses-1}
\end{eqnarray}
from Eq.(\ref{Eq:masses_1-2-3}).  Since $( M_{ij})_+$ for $i,j$=$e,\mu,\tau$ have a common phase of $\theta/2$ as in Eq.(\ref{Eq:Notations}), three phases of $-2\beta_{1,2,3}$ are determined to be $\theta/2$, which can be removed by the appropriate rotation of neutrinos.  No Majorana CP violation is active. 

It can be argued that $M^{(1)}_\nu$ itself is equivalent with $M^{(1)}_\nu$ with $\theta=0$.  By using Eq.(\ref{Eq:Notations}) for $\omega_\pm$ written in terms of $\vert \omega_\pm\vert$, $\tan 2\theta_{12,13}$ can be transformed into
\begin{eqnarray}
&&
\tan 2\theta _{12}  =
2\sqrt 2 \frac{\cos 2\theta _{13} \vert \left( M_{e\mu} \right)_+\vert}{c_{13} \left[ \left( 1-3s^2_{13}\right) \vert\left( M_{\mu\mu}\right)_+\vert - c_{13}^2 \left(\sigma\vert\left(M_{\mu\tau}\right)_+\vert + \vert\left(M_{ee}\right)_+\vert  \right)\right]},
\label{Eq:Angle12-1-theta=0}\\
&&
\tan 2\theta _{13} e^{i\delta } = 2\sqrt 2 \sigma \frac{i\vert\left( M_{e\mu} \right)_-\vert}{\vert\left( M_{\mu\mu} \right)_+\vert + \sigma \vert\left( M_{\mu\tau}\right)_+\vert + \vert\left( M_{ee}\right)_+\vert}.
\label{Eq:Angle13-1-theta=0}
\end{eqnarray}
These are equivalent to the expressions with $\vert (M_{ee,e\mu,\mu\mu,\mu\tau})_+\vert = {\rm Re}(M_{ee,e\mu,\mu\mu,\mu\tau})$ and $\vert (M_{e\mu})_-\vert = {\rm Im}(M_{e\mu})$.  Since all $\theta$-dependence can be removed in mixing angles and masses, $M^{(1)}_\nu$ itself is considered to be equivalent with $M^{(1)}_\nu$ with $\theta=0$.  Because of $\omega=e^{i\theta/2}(\vert\omega_+\vert + i\vert\omega_-\vert)$ for any complex values of $\omega$, we obtain that $M^{(1)}_\nu=e^{i\theta/2}(\vert M^{(1)}_{\nu +}\vert + i\vert M^{(1)}_{\nu -}\vert )$, where $\vert M^{(1)}_{\nu \pm}\vert$ is a mass matrix with $\vert (M_{ij})_\pm\vert$ in its element.  Our observation indicates that, after $\theta/2$ is rotated away, $M^{(1)}_\nu$ becomes $\vert M^{(1)}_{\nu +}\vert + i\vert M^{(1)}_{\nu -}\vert$, which only contributes to the non-vanishing $\delta$.  Therefore, $M^{(1)}_\nu$ with $\theta=0$ is a general form of the texture  $M^{(1)}_\nu$.

For $M^{(2)}_\nu$, 
\begin{eqnarray}
&&
X =  \frac{\sqrt 2\left( M_{e\mu} \right)_+}{c_{13}},
\quad
Y = \sqrt 2 \sigma \left( M_{e\mu} \right)_-,
\nonumber\\
&& 
\lambda _1  = \frac{s_{13}^2}{\cos 2\theta_{13}}\left( M_{\mu\mu} \right)_+
+M_{ee},
\quad
\lambda _2  = \left( M_{\mu\mu} \right)_+ + M_{ee},
\quad
\lambda _3  = \left( M_{\mu\mu} \right)_+ - M_{ee},
\label{Eq:Parameters-2}
\end{eqnarray}
are given. The masses and mixing angles are calculated to be:
\begin{eqnarray}
&& 
\tan 2\theta _{12}  =
2\sqrt 2 \frac{\cos 2\theta _{13} \left( M_{e\mu} \right)_+}{c_{13} \left( 1-3s^2_{13}\right) \left( M_{\mu\mu}\right)_+},
\quad
\tan 2\theta _{13} e^{i\delta } = 2\sqrt 2 \sigma \frac{\left( M_{e\mu} \right)_-}{\left( M_{\mu\mu} \right)_+},
\label{Eq:Angles-2}
\end{eqnarray}
and
\begin{eqnarray}
&& 
m_1 e^{ - 2i\beta _1 }  = 
\left( M_{\mu\mu} \right)_+ - \frac{1 + \cos 2\theta _{12} }{\sin 2\theta _{12} }\frac{\sqrt 2 \left( M_{e\mu} \right)_+}{c_{13} }+M_{ee},
\nonumber\\
&& 
m_2 e^{ - 2i\beta _2 }  = 
\left( M_{\mu\mu} \right)_+ + \frac{1 - \cos 2\theta _{12} }{\sin 2\theta _{12} }\frac{\sqrt 2 \left( M_{e\mu} \right)_+}{c_{13} }+M_{ee},
\nonumber\\
&& 
m_3 e^{ - 2i\beta _3 }  = \frac{c^2_{13}\left(  M_{\mu\mu} \right)_+}{\cos 2\theta _{13}}-M_{ee},
\label{Eq:Masses-2}
\end{eqnarray}
from Eq.(\ref{Eq:masses_1-2-3}). The three Majorana phases can take any values in this texture and the Majorana CP violation is active if ${\rm arg}(M_{ee})\neq\theta/2$.

For $M^{(3)}_\nu$, the results can be obtained by the replacement of $(M_{e\tau})_\pm\rightarrow (1\pm z)M_{e\mu}/2$ and $(M_{\mu\mu})_+\rightarrow (z\kappa+\kappa^\ast)M_{e\mu}/2$ in $M^{(2)}_\nu$.  The mixing angles are given by
\begin{eqnarray}
&& 
\tan 2\theta _{12}  =
2\sqrt 2 \frac{\cos 2\theta _{13} \left( 1+z \right)}{c_{13} \left( 1-3s^2_{13}\right) \left( z\kappa + \kappa^\ast\right)},
\quad
\tan 2\theta _{13} e^{i\delta } = 2\sqrt 2 \sigma\frac{1-z}{ z\kappa +\kappa^\ast},
\label{Eq:Angles-3}
\end{eqnarray}
which become consistent relations owing to Eq.(\ref{Eq:Notations2}). Masses are given by Eq.(\ref{Eq:Masses-2}) with the appropriate replacement.  As a result, three Majorana phases depend on phases of $M_{e\mu}$ and $M_{ee}$ and the Majorana CP violation is active if ${\rm arg}(M_{ee})\neq{\rm arg}(M_{e\mu})+\theta/2$ because $(1+ z)M_{e\mu}=\pm e^{{\rm arg}(M_{e\mu})+i\theta/2}\vert (1+ z)M_{e\mu}\vert$ and $(z\kappa+\kappa^\ast)M_{e\mu} = \pm e^{{\rm arg}(M_{e\mu})+i\theta/2}\vert(z\kappa+\kappa^\ast) M_{e\mu}\vert$.

\section{\label{sec:3}Observed Neutrino Properties}
To see the compatibility of the textures with the observed properties of neutrino oscillations, we examine whether $\sin^22\theta_{12}\gg\sin^2\theta_{13}$ \cite{Newest} as well as $\Delta m^2_{atm}\gg \Delta m^2_\odot$ is realized or not.  In the following discussions, we require $\sqrt{2}\mapleq\vert\tan 2\theta_{12}\vert\mapleq 2\sqrt{2}$ equivalent to $2/3\mapleq \sin^22\theta_{12}\mapleq 8/9$ and $\vert\tan 2\theta_{13}\vert\mapleq 1/2$ equivalent to $\sin^2\theta_{13}\mapleq 0.05$.  It is convenient to express $\tan 2\theta_{12}$ in terms of $x$ as $\tan 2\theta_{12} = 2\sqrt{2}/x$, giving $\sin^22\theta_{12}=8/(8+x^2)$ and $1\mapleq x\mapleq 2$.

For $M^{(1)}_\nu$ with $\theta=0$, the mixing angles are given by
\begin{eqnarray}
\tan 2\theta _{12}  &\approx& 2\sqrt 2 \frac{{\rm Re}\left( M_{e\mu} \right)}{ {\rm Re}\left( M_{\mu\mu} \right) - \sigma {\rm Re}\left( M_{\mu\tau}\right) - {\rm Re}\left( M_{ee}\right)},
\label{Eq:ApproxAngle12-1}\\
\tan 2\theta _{13} e^{i\delta } &=& 2\sqrt 2 \sigma \frac{i{\rm Im}\left( M_{e\mu} \right)}{{\rm Re}\left( M_{\mu\mu} \right) + \sigma {\rm Re}\left( M_{\mu\tau}\right) + {\rm Re}\left( M_{ee}\right)},
\label{Eq:ApproxAngle13-1}
\end{eqnarray}
where the expression of $\tan 2\theta _{12}$ is obtained by taking the approximation $\sin^2\theta_{13}\approx 0$ and this approximation is used in other textures, and the masses are
\begin{eqnarray}
&& 
m_1 e^{ - 2i\beta _1 }  \approx 
{\rm Re}\left( M_{\mu\mu} \right)-\sigma {\rm Re}\left( M_{\mu\tau} \right)- \frac{1 + \cos 2\theta _{12} }{\sin 2\theta _{12} }\sqrt 2 {\rm Re}\left( M_{e\mu} \right),
\nonumber\\
&& 
m_2 e^{ - 2i\beta _2 }  \approx 
{\rm Re}\left( M_{\mu\mu} \right)-\sigma {\rm Re}\left( M_{\mu\tau} \right) + \frac{1 - \cos 2\theta _{12} }{\sin 2\theta _{12} }\sqrt 2 {\rm Re}\left( M_{e\mu} \right),
\nonumber\\
&& 
m_3 e^{ - 2i\beta _3 }  \approx {\rm Re}\left( M_{\mu\mu} \right) +\sigma {\rm Re}\left( M_{\mu\tau} \right).
\label{Eq:ApproxMasses-1}
\end{eqnarray}
It is evident that $\beta_{1,2,3}=0$ indicating no Majorana CP violation.  From Eqs.(\ref{Eq:ApproxAngle12-1}) and (\ref{Eq:ApproxAngle13-1}), we find that the condition of $\sqrt{2}\mapleq\vert\tan 2\theta_{12}\vert\mapleq 2\sqrt{2}$ and $\vert\tan 2\theta_{13}\vert \mapleq 1/2$ is satisfied by 
\begin{eqnarray}
&&
2 \left| {\rm Re}\left( M_{e\mu}\right)\right| \mapgeq \left| {\rm Re}\left( M_{\mu\mu} \right) - \sigma {\rm Re}\left( M_{\mu\tau}\right) - \left( {\rm Re}M_{ee}\right)\right|\mapgeq\left| {\rm Re}\left( M_{e\mu}\right)\right|,
\nonumber \\
&&
\left| {\rm Re}\left( M_{\mu\mu} \right) + \sigma {\rm Re}\left( M_{\mu\tau}\right) + {\rm Re}\left( M_{ee}\right)\right|
\mapgeq
4\sqrt{2}\left| {\rm Im} \left( M_{e\mu} \right)\right|.
\label{Eq:Condition12-13-1}
\end{eqnarray}
It turns out that there are sufficient freedoms to explain the mass hierarchy of $\Delta m^2_{atm}\gg \Delta m^2_\odot$ while Eq.(\ref{Eq:Condition12-13-1}) is satisfied.

For $M^{(2)}_\nu$, the mixing angles are given by
\begin{eqnarray}
&& 
\tan 2\theta _{12}  \approx
2\sqrt 2 \frac{\left( M_{e\mu} \right)_+}{\left( M_{\mu\mu}\right)_+},
\quad
\tan 2\theta _{13} e^{i\delta } = 2\sqrt 2 \sigma \frac{\left( M_{e\mu} \right)_-}{\left( M_{\mu\mu} \right)_+}.
\label{Eq:ApproxAngles-2}
\end{eqnarray}
The masses are
\begin{eqnarray}
&& 
m_1 e^{ - 2i\beta _1 }  \approx 
\left( M_{\mu\mu} \right)_+- \frac{1 + \cos 2\theta _{12} }{\sin 2\theta _{12} }\sqrt 2 \left( M_{e\mu} \right)_++M_{ee},
\nonumber\\
&& 
m_2 e^{ - 2i\beta _2 }  \approx 
\left( M_{\mu\mu} \right)_+ + \frac{1 - \cos 2\theta _{12} }{\sin 2\theta _{12} }\sqrt 2 \left( M_{e\mu} \right)_++M_{ee},
\nonumber\\
&& 
m_3 e^{ - 2i\beta _3 }  \approx \left( M_{\mu\mu} \right)_+ -M_{ee} ,
\label{Eq:ApproxMasses-2}
\end{eqnarray}
which are further transformed into
\begin{eqnarray}
m_1 e^{ - 2i\beta _1 }  &\approx& - \frac{\sqrt {8 + x^2 }  - x}{2}\left( M_{e\mu} \right)_+ + M_{ee},
\quad
m_2 e^{ - 2i\beta _2 }  \approx     \frac{\sqrt {8 + x^2 }  + x}{2}\left( M_{e\mu} \right)_+ + M_{ee},
\nonumber\\
m_3 e^{ - 2i\beta _3 }  &\approx&   x \left( M_{e\mu} \right)_+- M_{ee},
\label{Eq:ApproxMasses2-2}
\end{eqnarray}
where $x = (M_{\mu\mu})_+/(M_{e\mu})_+$.

From Eq.(\ref{Eq:ApproxAngles-2}), we find that the condition of $\sqrt{2}\mapleq\vert\tan 2\theta_{12}\vert\mapleq 2\sqrt{2}$ and $\vert\tan 2\theta_{13}\vert \mapleq 1/2$ is satisfied by 
\begin{eqnarray}
&&
2 \left| \left( M_{e\mu}\right)_+\right| \mapgeq \left| \left( M_{\mu\mu} \right)_+\right|
\mapgeq
\left| \left( M_{e\mu} \right)_+\right|,
\quad
\left| \left( M_{\mu\mu} \right)_+\right|
\mapgeq
4\sqrt{2}\left| \left( M_{e\mu} \right)_-\right|.
\label{Eq:Condition12-13-2}
\end{eqnarray}
Because
\begin{eqnarray}
\Delta m_ \odot ^2  &=&
 \sqrt {8 + x^2 } \left[\left( 2{\rm Re} \left( M_{ee} \right) + x{\rm Re} \left( M_{e\mu},\theta \right) \right){\rm Re} \left( M_{e\mu},\theta \right)+\left( 2{\rm Im} \left( M_{ee} \right) + x{\rm Im} \left( M_{e\mu},\theta \right) \right){\rm Im} \left( M_{e\mu},\theta \right)\right],
\nonumber\\
\Delta m_{atm}^2  &=& \frac{\sqrt {8 + x^2 }  + 3x}{4}\left| \left[ 4{\rm Re} \left( M_{ee} \right) + \left( \sqrt {8 + x^2 }  - x \right){\rm Re} \left( M_{e\mu},\theta \right) \right]{\rm Re} \left( M_{e\mu},\theta \right)\right.
\nonumber\\
&&
\left.+
\left[ 4{\rm Im} \left( M_{ee} \right) + \left( \sqrt {8 + x^2 }  - x \right){\rm Im} \left( M_{e\mu},\theta \right) \right]{\rm Im} \left( M_{e\mu},\theta \right)  \right|,
\label{Eq:MassSquared-2}
\end{eqnarray}
the hierarchy of $\Delta m^2_{atm}\gg\Delta m^2_\odot$ can be realized by requiring that 
\begin{eqnarray}
\left( {2{\rm Re} \left( M_{ee} \right) + x{\rm Re} \left( M_{e\mu},\theta  \right)} \right){\rm Re} \left( M_{e\mu},\theta \right) + \left( {2{\rm Im} \left( M_{ee} \right) + x{\rm Im} \left( M_{e\mu},\theta \right)} \right){\rm Im} \left( M_{e\mu},\theta  \right) \approx 0.
\label{Eq:ConditionMass-2}
\end{eqnarray}
Three neutrino masses are, then, approximated to be:
\begin{eqnarray}
m_1^2  &\approx& m_2^2 \approx 
\frac{8 + x^2 }{4}\left({\rm Re}\left( M_{e\mu},\theta \right)^2+{\rm Im}\left( M_{e\mu},\theta \right)^2 \right)
\nonumber\\
&&
+
\frac{1}{4}\left[
\left(2{\rm Re} \left( M_{ee} \right)+x{\rm Re}\left( M_{e\mu},\theta \right)\right)^2
+
\left(2{\rm Im} \left( M_{ee} \right)+x{\rm Im}\left( M_{e\mu},\theta \right)\right)^2
\right],
\nonumber\\
m_3^2  &\approx& \frac{9x^2 }{4}\left({\rm Re}\left( M_{e\mu},\theta \right)^2+{\rm Im}\left( M_{e\mu},\theta \right)^2 \right)
\nonumber\\
&&
+
\frac{1}{4}\left[
\left(2{\rm Re} \left( M_{ee} \right)+x{\rm Re}\left( M_{e\mu},\theta \right)\right)^2
+
\left(2{\rm Im} \left( M_{ee} \right)+x{\rm Im}\left( M_{e\mu},\theta \right)\right)^2
\right],
\label{Eq:DegenerateMass-2}
\end{eqnarray}
which are masses of degenerate neutrinos because of $x={\mathcal{O}}$(1).   As a result, $\Delta m_{atm}^2$ becomes
\begin{eqnarray}
\Delta m_{atm}^2  \approx 2\left( {x^2  - 1} \right)\left( {{\rm Re} \left( {M_{e\mu},\theta } \right)^2  + {\rm Im} \left( {M_{e\mu},\theta } \right)^2 } \right),
\label{Eq:ApproxAtmMass-2}
\end{eqnarray}
indicating that
\begin{eqnarray}
{{\rm Re} \left( {M_{e\mu},\theta } \right)^2  + {\rm Im} \left( {M_{e\mu},\theta } \right)^2 } \approx {\mathcal{O}}(10^{-3})~{\rm eV}^2.
\label{Eq:ApproxAtmMass-2-1}
\end{eqnarray}
The size of the neutrino masses varies with the common magnitude of $(2{\rm Re} ( M_{ee})+x{\rm Re}( M_{e\mu},\theta))^2+(2{\rm Im} ( M_{ee})+x{\rm Im}( M_{e\mu},\theta))^2$.

For $M^{(3)}_\nu$ with the appropriate replacement in $M^{(2)}_\nu$, the mixing angles are given by
\begin{eqnarray}
&& 
\tan 2\theta _{12} \approx 
2\sqrt 2 \frac{1+z}{z\kappa+\kappa^\ast},
\quad
\tan 2\theta _{13} e^{i\delta } = 2\sqrt 2 i \sigma\frac{1-z}{z\kappa+\kappa^\ast}.
\label{Eq:ApproxAngles-3}
\end{eqnarray}
The similar discussion to obtain the hierarchy of $\Delta m^2_{atm}\gg\Delta m^2_\odot$ yields
\begin{eqnarray}
&&
\left( {4{\rm Re} \left( M_{ee} \right) + x{\rm Re} \left( (1+z)M_{e\mu}\right)} \right){\rm Re} \left( (1+z)M_{e\mu} \right) 
+
 \left( {4{\rm Im} \left( M_{ee} \right) + x{\rm Im} \left( (1+z)M_{e\mu}\right)} \right){\rm Im} \left( (1+z)M_{e\mu}\right) \approx 0,
\label{Eq:ConditionMass-3}
\end{eqnarray}
and $\Delta m_{atm}^2$ is given by 
\begin{eqnarray}
\Delta m_{atm}^2  \approx \frac{x^2  - 1}{2}\left| (1+z)M_{e\mu} \right|^2,
\label{Eq:MassSquared-3}
\end{eqnarray}
where $x = (z\kappa+\kappa^\ast)/(1+z)$.   Corresponding to Eq.(\ref{Eq:Condition12-13-2}), we obtain that
\begin{eqnarray}
&&
2 \left| 1+z\right| \mapgeq \left| z\kappa+\kappa^\ast\right|
\mapgeq
\left| 1+z\right|,
\quad
\left| z\kappa+\kappa^\ast\right|
\mapgeq
4\sqrt{2}\left| 1-z \right|.
\label{Eq:Condition12-13-3}
\end{eqnarray}
The size of $\left| M_{e\mu} \right|$ is of order of $\sqrt{\Delta m_{atm}^2}$.

\section{\label{sec:4}Summary and Discussions}
Examining the condition of $\tan\theta_{23}={\rm Im}( {\rm\bf M}_{e\mu})/{\rm Im}( {\rm\bf M}_{e\tau})$  for ${\rm\bf M}=M^\dagger_\nu M_\nu$ satisfied by any textures, we have found three types of textures of $M_\nu$ that provide maximal CP violation and maximal atmospheric neutrino mixing.  The key condition on the appearance of maximal CP violation is $\sin 2\theta_{23}{\rm\bf M}_{e\tau}=-{\rm\bf M}^\ast_{e\mu}+\cos 2\theta_{23}{\rm\bf M}_{e\mu}$, yielding Eq.(\ref{Eq:MaximalAtm-CP-M}) for maximal atmospheric neutrino mixing.  The three types of textures suggested by Eq.(\ref{Eq:MaximalAtm-CP-M}), which are controlled by one phase of $\theta$ as $z=e^{i\theta}$, are described by
\begin{eqnarray}
&& 
	M^{(1)}_\nu = \left( {\begin{array}{*{20}c}
   M_{ee} & M_{e\mu} & -\sigma zM^\ast_{e\mu} \\
   M_{e\mu} & M_{\mu\mu} & M_{\mu\tau}  \\
   -\sigma zM^\ast_{e\mu} & M_{\mu\tau} & zM^\ast_{\mu\mu}  \\
\end{array}} \right),
\quad
	M^{(2)}_\nu = \left( {\begin{array}{*{20}c}
   M_{ee} & M_{e\mu} & -\sigma zM^\ast_{e\mu} \\
   M_{e\mu} & M_{\mu\mu} & -\sigma M_{ee}  \\
   -\sigma zM^\ast_{e\mu} & -\sigma M_{ee} & zM^\ast_{\mu\mu}  \\
\end{array}} \right),
\nonumber\\
&& 
	M^{(3)}_\nu = \left( {\begin{array}{*{20}c}
   M_{ee} & M_{e\mu} & -\sigma zM_{e\mu} \\
   M_{e\mu} & z\kappa M_{e\mu} & -\sigma M_{ee}  \\
   -\sigma zM_{e\mu} & -\sigma M_{ee} &  \kappa^\ast M_{e\mu}  \\
\end{array}} \right),
\label{Eq:Mnu-Summary}
\end{eqnarray}
where $M_{ee,\mu\tau}=(M_{ee,\mu\tau})_+$ is imposed on $M^{(1)}_\nu$. It is found that $M^{(1)}_\nu$ with $\theta=0$ represents a general texture of this type.  The three Majorana phases are calculable in principle and shows that the Majorana CP violation is 
\begin{itemize}
\item absent for $M^{(1)}_\nu$,
\item present if ${\rm arg}(M_{ee})\neq\theta/2$ for $M^{(2)}_\nu$,
\item present if ${\rm arg}(M_{ee})\neq{\rm arg}(M_{e\mu})+\theta/2$ for $M^{(3)}_\nu$.
\end{itemize}
To obtain these specific textures depends on how the CP violating Dirac phase of $\delta$ is parameterized in $U_{PMNS}$.  For instance, another parameterization of the Kobayashi-Maskawa type \cite{K-M} gives $\tan\theta_{23}=-{\rm Im}({\rm\bf M}_{e\tau})/{\rm Im}({\rm\bf M}_{e\mu})$ for Eq.(\ref{Eq:theta23}), $c_{23} {\rm\bf M}_{e\mu} - s_{23} {\rm\bf M}_{e\tau} = \left| {c_{23} {\rm\bf M}_{e\mu} - s_{23} {\rm\bf M}_{e\tau}} \right| e^{-i\delta }$ for Eq.(\ref{Eq:Phase-delta}) and  $\tan\theta_{23}={\rm Re}({\rm\bf M}_{e\mu})/{\rm Re}({\rm\bf M}_{e\tau})$ for Eq.(\ref{Eq:MaximalCPRelation}) \cite{GeneralCP}. As a result, Eq.(\ref{Eq:MaximalAtm-CP}) is replaced by ${\rm\bf M}_{e\tau}=\sigma {\rm\bf M}^\ast_{e\mu}$.  However, it is obvious that, for the obtained textures, the diagonalization is only possible by $U_{PMNS}$ of Eq.(\ref{Eq:U_nu}).  Namely, if we have a given texture, then the form of $U_{PMNS}$ is completely determined \cite{Koide}.

The mixing angles of $\theta_{12,13}$ are given by
\begin{eqnarray}
\tan 2\theta _{12}  &\approx& 2\sqrt 2 \frac{\left( M_{e\mu} \right)_+}{\left( M_{\mu\mu} \right)_+ - \sigma \left( M_{\mu\tau}\right)_+ - \left( M_{ee}\right)_+},
\nonumber\\
\tan 2\theta _{13} e^{i\delta } &=& 2\sqrt 2 \sigma \frac{\left( M_{e\mu} \right)_-}{\left( M_{\mu\mu} \right)_+ + \sigma \left( M_{\mu\tau}\right)_+ + \left( M_{ee}\right)_+},
\label{Eq:ApproxAnglesSummary-1}
\end{eqnarray}
for $M^{(1)}_\nu$, where $(M_{ij})_+$ and $(M_{ij})_-$ ($i,j$=$e,\mu,\tau$) are, respectively, reduced to ${\rm Re}(M_{ij})$ and $i{\rm Im}(M_{ij})$ at $\theta=0$, by
\begin{eqnarray}
&& 
\tan 2\theta _{12} \approx 2\sqrt 2 \frac{\left( M_{e\mu} \right)_+}{\left( M_{\mu\mu} \right)_+},
\quad
\tan 2\theta _{13} e^{i\delta } = 2\sqrt 2 \sigma \frac{\left( M_{e\mu} \right)_-}{\left( M_{\mu\mu} \right)_+},
\label{Eq:ApproxAnglesSummary-2}
\end{eqnarray}
for $M^{(2)}_\nu$, and by
\begin{eqnarray}
&& 
\tan 2\theta _{12} \approx 
2\sqrt 2 \frac{1+z}{z\kappa+\kappa^\ast},
\quad
\tan 2\theta _{13} e^{i\delta } = 2\sqrt 2 i \sigma\frac{1-z}{z\kappa+\kappa^\ast}.
\label{Eq:ApproxAnglesSummary-3}
\end{eqnarray}
for $M^{(3)}_\nu$, where $\delta= \pm \pi/2$.  The pure imaginary value of $e^{i\delta }$ ensured by Eq.(\ref{Eq:MaximalAtm-CP-M}) is assured by the fact that  $(\omega^\prime )_-/(\omega )_+$ and $(1-z)/(z\omega+\omega^\ast)$ become pure imaginary for any complex values of $\omega$ and $\omega^\prime$.  We have then demonstrated that there are indeed consistent flavor neutrino masses that reproduce $\sin^22\theta_{12}\gg\sin^2\theta_{13}\approx 0$ and $\Delta m^2_{atm}\gg\Delta m^2_\odot$. For $M^{(2,3)}_\nu$, neutrinos turn out be degenerate ones. It is further recognized that $M^{(1,2)}_\nu$ exhibit a novel feature that $\tan 2\theta_{12}$ is proportional to $(M_{e\mu})_+$ while $\tan 2\theta_{13}$ is proportional to $(M_{e\mu})_-$.  It can be stated in the simplest case of $z=\pm 1$ that
\begin{itemize}
\item $\tan 2\theta_{12}$ is proportional to Re($M_{e\mu}$) for $z=1$ and Im($M_{e\mu}$) for $z=-1$,
\item $\tan 2\theta_{13}$ is proportional to Im($M_{e\mu}$) for $z=1$ and Re($M_{e\mu}$) for $z=-1$.
\end{itemize}
This feature is absence in models without CP violation.

Although we have not specified a mechanism to create flavor neutrino masses, it is conceivable that these masses are induced by the seesaw mechanism \cite{Seesaw} and that the specific structure in phases presented in Eq.(\ref{Eq:Mnu-Summary}) may be realized at the seesaw scale of ${\mathcal {O}}(10^{12}-10^{14})$ GeV.  If this is the case, radiative corrections to yield renormalized values around the weak scale corresponding to the observed values may significantly alter our phase structure, especially the one giving $\delta = \pm \pi/2$, in the textures.  However, it is known that such radiative corrections are small in the normal mass hierarchy case \cite{Normal}. In the inverted mass hierarchy case and near degeneracy case, sizable corrections may arise in certain parameter regions \cite{InvertedDegenerated}.  The form of $M^{(1)}$ in the normal mass hierarchy case is, thus, stable against radiative corrections.  On the other hand, $M^{(2,3)}$ leading to the near degeneracy case may receive large radiative corrections, which have to be estimated in a specific model that yields $M^{(2,3)}$ via the seesaw mechanism.  Conversely, one can extrapolate the form of $M^{(2,3)}$ at the seesaw scale that indeed yields $M^{(2,3)}$ in Eq.(\ref{Eq:Mnu-Summary}) at the weak scale.

\appendix
\section{\label{sec:Appendix}Neutrino Masses and Mixing Angles}

After the direct calculation of $U^T_{PMNS}M_\nu U_{PMNS}$, one can obtain constraints:
\begin{eqnarray}
&&
c_{12} \Delta_1  - s_{12} c_{13} \left( {s_{13} e^{ - i\delta } X + \Delta_2 } \right) = 0,
\quad
s_{12} \Delta_1  + c_{12} c_{13} \left( {s_{13} e^{ - i\delta } X + \Delta_2 } \right) = 0,
\label{Eq:Constraint1} \\
&&
c_{12} \left[ {s_{12} \lambda_1  + c_{12} \left( {c_{13}^2 X - s_{13} e^{i\delta } \Delta_2 } \right)} \right] - s_{12} \left[ {c_{12} \lambda_2  + s_{12} \left( {c_{13}^2 X - s_{13} e^{i\delta } \Delta_2 } \right)} \right] = 0,
\label{Eq:Constraint2}
\end{eqnarray}
where
\begin{eqnarray}
&&
\Delta_1  = \frac{{M_{ee}e^{ - i\delta }  - \lambda_3 e^{i\delta } }}{2}\sin 2\theta_{13}  + Y\cos 2\theta_{13} ,
\quad
\Delta_2  = M_{\mu\tau}\cos 2\theta_{23}  - \frac{M_{\tau\tau} - M_{\mu\mu}}{2}\sin 2\theta_{23} ,
\label{Eq:Delta1-Delta2}\\
&&
X = \frac{c_{23} M_{e\mu} - s_{23} M_{e\tau}}{c_{13}},
\quad
Y = s_{23} M_{e\mu} + c_{23} M_{e\tau},
\label{Eq:X-Y}
\end{eqnarray}
and diagonalized masses:
\begin{eqnarray}
&&
m_1 e^{ - 2i\beta _1 }  = c_{12}^2 \lambda_1  + s_{12}^2 \lambda _2  - 2c_{12} s_{12} X,
\quad
m_2 e^{ - 2i\beta _2 }  = s_{12}^2 \lambda_1  + c_{12}^2 \lambda _2  + 2c_{12} s_{12} X,
\nonumber\\
&&
m_3 e^{ - 2i\beta _3 }  = c_{13}^2 \lambda _3  + 2c_{13} s_{13} e^{ - i\delta } Y + s_{13}^2 e^{ - 2i\delta } M_{ee},
\label{Eq:Masses}
\end{eqnarray}
where
\begin{eqnarray}
&&
\lambda_1  = c_{13}^2 M_{ee} - 2c_{13} s_{13} e^{i\delta } Y + s_{13}^2 e^{2i\delta }\lambda_3,
\quad
\lambda_2  = c_{23}^2 M_{\mu\mu} + s_{23}^2 M_{\tau\tau} - 2s_{23} c_{23} M_{\mu\tau},
\nonumber\\
&&
\lambda_3  = s_{23}^2 M_{\mu\mu} + c_{23}^2 M_{\tau\tau} + 2s_{23} c_{23} M_{\mu\tau}.
\label{Eq:Parameters}
\end{eqnarray}
From Eq.(\ref{Eq:Constraint2}) with $s_{13} e^{ - i\delta } X + \Delta_2=0$ in Eq.(\ref{Eq:Constraint1}), $\theta_{12}$ is determined by
\begin{eqnarray}
&&
{\sin 2\theta _{12} \left( {\lambda _1  - \lambda _2 } \right) + 2\cos 2\theta _{12} X = 0},
\label{Eq:theta12-Appendix}
\end{eqnarray}
and, from $\Delta_1=0$ in Eq.(\ref{Eq:Constraint1}), $\theta_{13}$ is determined by
\begin{eqnarray}
&&
{\sin 2\theta _{13} \left( {M_{ee}e^{ - i\delta }  - \lambda _3 e^{i\delta } } \right) + 2\cos 2\theta _{13} Y} = 0.
\label{Eq:theta13-Appendix}
\end{eqnarray}
Also from Eq.(\ref{Eq:Constraint1}), the atmospheric mixing angle is given by
\begin{eqnarray}
&&
\left( M_{\tau\tau} - M_{\mu\mu}\right)\sin 2\theta_{23}  - 2 M_{\mu\tau}\cos 2\theta_{23}= 2s_{13} e^{ - i\delta } X.
\label{Eq:theta23-Appendix}
\end{eqnarray}
Eqs.(\ref{Eq:theta12-Appendix}) and (\ref{Eq:theta13-Appendix}) further convert the mass parameters into
\begin{eqnarray}
&&
m_1 e^{ - 2i\beta_1 }  = \frac{{\lambda_1  + \lambda_2 }}{2} - \frac{X}{{\sin 2\theta_{12} }},
\quad
m_2 e^{ - 2i\beta_2 }  = \frac{{\lambda_1  + \lambda_2 }}{2} + \frac{X}{{\sin 2\theta_{12} }},
\nonumber\\
&&
m_3 e^{ - 2i\beta_3 }  =\frac{c_{13}^2 \lambda _3 - s_{13}^2 e^{-2i\delta } M_{ee} }{\cos 2\theta _{13} },
\label{Eq:masses_1-2-3}
\end{eqnarray}
and
\begin{eqnarray}
&&
\Delta m_ \odot ^2  = \frac{2}{\sin 2\theta _{12} }{\rm Re} \left( {\left( \lambda^ \ast _1  + \lambda^ \ast _2 \right)  X} \right).
\label{Eq:SolarSquared}
\end{eqnarray}

\end{document}